\documentclass[11pt]{article}

\usepackage{graphicx}%
\usepackage{multirow}%
\usepackage{amsmath,amssymb,amsfonts}%
\usepackage{amsthm}%
\usepackage{mathrsfs}%
\usepackage[title]{appendix}%
\usepackage{xcolor}%
\usepackage{textcomp}%
\usepackage{manyfoot}%
\usepackage{booktabs}%
\usepackage{algorithm}%
\usepackage{algorithmicx}%
\usepackage{algpseudocode}%
\usepackage{listings}%

\usepackage{dcolumn}
\usepackage{bm,mathtools}
\usepackage{cancel}
\usepackage{soul}
\usepackage[normalem]{ulem}
\usepackage{physics}

\usepackage[english]{babel}
\usepackage[a4paper,margin=2cm]{geometry}

\usepackage{tikz}
\usepackage{pgfplots}

\usepackage{enumitem}
\usepackage{float}
\usepackage{caption}
\usepackage{multicol}
\usepackage{parskip}
\usepackage{comment}
\usepackage{microtype}
\usepackage{natbib}
\usepackage{url}

\DeclareMathOperator{\atanh}{atanh}

\usepackage{thmtools}

\theoremstyle{plain}%
\newtheorem{theorem}{Theorem}
\newtheorem{proposition}[theorem]{Proposition}%
\newtheorem{lemma}[theorem]{Lemma}%
\newtheorem{corollary}[theorem]{Corollary}%

\theoremstyle{remark}%

\theoremstyle{definition}%
\newtheorem{definition}{Definition}%
\renewenvironment{proof}[1][Proof]{\par\noindent\textbf{#1.}\ }{\hfill$\square$\par}
\DeclareMathOperator{\Span}{span}

\title{Oscillatory instability of subcritical stationary solutions in the parametrically driven, damped nonlinear Dirac equation}

\author{Fernando Carre\~no-Navas\\
\small IMUS and Departamento de An\'alisis Matem\'atico, Universidad de Sevilla, c/Tarfia s/n, Sevilla, 41012, Spain\\
Email: \texttt{fcarreno@us.es}
}

\date{19 June 2026}

\begin{document}
\maketitle

\begin{abstract}
The dynamics of the nonlinear Dirac equation (NLDE) under the influence of dissipation and parametric forcing exhibit complex phenomenologies, including soliton destruction and chaotic transitions, which have been extensively observed in numerical simulations. However, a rigorous analytical understanding of the underlying instability mechanisms in the subcritical regime has remained open.

In this work, we provide a theoretical answer to these numerical findings by establishing the spectral instability of the fundamental soliton solution. Our methodology relies on a three-step approach linking local asymptotic analyses with global topological arguments. First, we characterize the generalized null space at the critical parameter ($\lambda_c$), revealing the onset of instability through the splitting of the zero eigenvalue within a parity-dependent subspace. Second, in the non-relativistic limit ($\lambda \to 0$), we use a Lyapunov--Schmidt reduction to reduce the linearized Dirac's operators into a P\"oschl--Teller Schr\"odinger operator, thereby identifying the instability source as a unique real pair of discrete eigenvalues in the restricted problem. Third, we extend these local results to the entire subcritical domain using a topological continuation argument, which tracks the invariance of the spectral parity in a subset of the complex plane. These analytical results rigorously confirm the instability of the soliton, providing the mathematical mechanism for the wave collapse observed in the driven-damped system.

\end{abstract}


\section{Introduction}

The Nonlinear Dirac Equation (NLDE) has emerged as a fundamental model in mathematical physics, extending well beyond its original scope in relativistic quantum field
theory \citep{strauss1986}. It provides deep insights into systems where nonlinearity and dispersion compete, serving as an effective description for phenomena ranging
from gap solitons in photonic lattices and optical fibers to charge transport in conducting polymers and Bose-Einstein condensates in honeycomb lattices \citep{aceves1989,campbell1982, kivshar2003, grossneveu1974}.

In experimental settings, these systems are rarely isolated; they are inevitably subject to environmental damping, which causes the amplitude of
solitary waves to decay. Consequently, an active area of research is the stabilization of these structures via external energy injection. The application of a parametric
driving force offers an effective mechanism to compensate for such dissipative losses, allowing for the existence of sustained, steady soliton states. Understanding and
controlling these relativistic solitary waves is crucial for developing robust technologies at the frontier of nonlinear optics and condensed matter physics.

In this paper, we focus on a modification of the Gross-Neveu model of the NLDE, recently proposed in \citep{chaos_nlde}
\begin{equation}\label{NLDE}
	i \gamma^\mu \partial_\mu \Psi - \Psi + (\bar{\Psi}\Psi) \Psi = r e^{-i2 \Omega t} \Psi^\star - i\rho \gamma^0 \Psi,
\end{equation}
where $r>0$ represents the amplitude of the parametric force, $2\Omega>0$ is the driving frequency, and $\rho>0$ is the dissipation coefficient. The Dirac adjoint is defined as $\bar{\Psi} = (\Psi^\star)^T \gamma^0$, and we adopt the standard representation for the Dirac matrices
\begin{equation*}
	\gamma^0 = \begin{pmatrix}
		1 & 0  \\
		0 & -1
	\end{pmatrix}, \qquad \gamma^1 = \begin{pmatrix}
		0 & 1 \\
		1 & 0
	\end{pmatrix}.
\end{equation*}

The addition of linear dissipation and parametric driving breaks the Lorentz invariance and the Hamiltonian structure of the classical equation,
introducing a rich variety of dynamical behaviors. Despite the loss of these structural symmetries, exact analytical solutions can still be found.

Prior investigations in \citep{chaos_nlde} revealed the existence of two distinct families of stationary soliton solutions, which we denote by $\Psi_+$ (supercritical) and $\Psi_-$ (subcritical), defined by
\begin{equation}\label{solitons}
	\Psi_\pm(x, t) = \begin{pmatrix}
		\psi_\pm(x) \\ i \chi_\pm(x)
	\end{pmatrix}e^{-i(\Omega t+ \Theta_\pm/2)}.
\end{equation}
The phase angles are determined by an exact balance between gain and loss, $\sin(\Theta_\pm) = \rho/r$, leading to two solutions $\Theta_+ = \arcsin(\rho/r)$ and $\Theta_- = \pi-\Theta_+$. The spatial profiles are defined as
\begin{equation}\label{def_phi}
	\psi_\pm(x) = \sqrt{2} \beta_\pm \frac{\sqrt{1+\omega_\pm} \sech(\beta_\pm x)}{2\omega_\pm+(1+\omega_\pm)\sech^2(\beta_\pm x)}, \qquad \chi_\pm(x) = \sqrt{\frac{1-\omega_\pm}{1+\omega_\pm}} \tanh(\beta_\pm x) \psi_\pm(x),
\end{equation}
where $\omega_\pm = \Omega \mp \sqrt{r^2-\rho^2}$ corresponds to the detuning of the soliton frequency, and $\beta_\pm = \sqrt{1-\omega_\pm^2}$ is the inverse width of the soliton. The existence of these solutions requires the condition $r > \rho$, $0 < \omega_\pm < 1$ and for the specific case of $\Psi_-$, $0<\Omega<1$.

Previous studies addressing the stability of these exact solutions have relied primarily on variational approaches (see \citep{Shao_2014})
and numerical computations of the eigenvalue spectrum \citep{chaos_nlde}. While variational methods map the infinite-dimensional dynamics onto a
finite-dimensional planar system, providing intuitive phase portraits, they impose rigid constraints on the perturbation profile and often fail
to rigorously capture complex spectral mechanisms across the entire parameter space. Recent numerical explorations using Chebyshev differentiation
have provided strong numerical evidence that the subcritical $\Psi_-$ soliton is unstable throughout its existence regime, while the supercritical $\Psi_+$ soliton possesses
a stability window. However, a rigorous analytical understanding of these spectral properties has remained an open problem.

Unlike the nonlinear Schr\"odinger equation (NLSE), where the linearized operator is bounded from below and well-established tools like Pelinovsky's eigenvalue-counting formula (see \citep{pelinovsky_periodic_potentials}) can be applied, the Dirac operator is strongly indefinite. Its spectrum is unbounded from both above and below, rendering standard variational arguments inapplicable and significantly complicating the identification of unstable eigenvalues emerging from the essential spectrum or the origin.

To analyze the spectral stability, we consider small perturbations around the exact stationary solutions. We introduce the linearization ansatz:
\begin{equation}\label{ansatz}
	\Psi(x,t) = \left[ \Phi_\pm(x) + \delta \Phi(x,t) \right] e^{-i(\Omega t + \Theta_\pm/2)},
\end{equation}
where $\Phi_\pm(x) = (\psi_\pm(x), i\chi_\pm(x))^T$ represents the unperturbed spatial profile. Substituting \eqref{ansatz} into Eq.~\eqref{NLDE} and retaining only linear terms in $\delta \Phi$ yields a set of coupled linear evolution equations.

Because the original system depends on three independent physical parameters $(\rho, r, \Omega)$, performing a direct spectral analysis is highly cumbersome. To overcome this, we utilize a coordinate and scaling transformation introduced in \citep{chaos_nlde}. Under this change of variables, the stability problem is cast into a unified non-self-adjoint eigenvalue problem:
\begin{equation}\label{sistema_inicial}
	L: H^1(\mathbb{R})^4 \mapsto L^2(\mathbb{R})^4, \qquad L y =
	\begin{pmatrix}
		0    & L_0 \\
		-L_1 & 0
	\end{pmatrix} y = -\Lambda y,
\end{equation}
where $\Lambda$ is the spectral parameter and the block operators $L_0$ and $L_1$ take the form
\begin{align}\label{def_operadores}
	L_0 & = \begin{pmatrix}
		        -\tilde{N} + \frac{1}{\lambda} - \frac{1+\lambda^2}{\lambda}\Omega & \partial_x                                              \\
		        -\partial_x                                                        & \tilde{N} - \lambda - \frac{1+\lambda^2}{\lambda}\Omega
	        \end{pmatrix}, \qquad
	L_1 = \begin{pmatrix}
		      -\tilde{N} - 2\psi^2 + \lambda & \partial_x + 2\psi\chi                  \\
		      -\partial_x + 2\psi\chi        & \tilde{N} - 2\chi^2 - \frac{1}{\lambda}
	      \end{pmatrix}.
\end{align}

The potential terms depend on the scaled auxiliary functions $\psi(x)$ and $\chi(x)$, and $\tilde{N}(x) = \psi^2(x) - \chi^2(x)$, explicitly given by:
\begin{equation}\label{soliton_profile}
	\psi(x) = \frac{\sqrt{2\lambda}\,\sech(x)}{(1-\lambda^2)+\lambda^2 \sech^2(x)}, \qquad \chi(x) = \lambda \tanh(x)\,\psi(x),
\end{equation}
which satisfy the following system:
\begin{equation}\label{sistema_psi_chi}
	\begin{split}
		\partial_x{\psi} + \frac{1}{\lambda}{\chi} - \bigl({\psi}^2 - {\chi}^2\bigr){\chi} = 0, \\
		\partial_x{\chi} + \lambda {\psi} - \bigl({\psi}^2 - {\chi}^2\bigr){\psi} = 0.
	\end{split}
\end{equation}

A crucial feature of this formulation is that the system can be expressed in terms of a single dimensionless parameter $\lambda$, defined by
\begin{equation*}
	\lambda = \sqrt{\frac{1-\omega_\pm}{1+\omega_\pm}}, \quad \lambda \in (0, 1),
\end{equation*}
which encodes the dependence on the physical parameters $\rho$ and $r$.

This unification elegantly captures both soliton families $\Psi_\pm$ within the interval $(0,1)$. The boundary between the two branches corresponds to a critical bifurcation point $\lambda_c$, explicitly given by:
\begin{equation*}
	\lambda_c = \sqrt{\frac{1-\Omega}{1+\Omega}}.
\end{equation*}

The subcritical regime $\lambda \in (0, \lambda_c)$ corresponds exclusively to the soliton $\Psi_-$, while the supercritical regime $\lambda \in (\lambda_c, 1)$ corresponds to $\Psi_+$.

Finally, the connection between the spectrum $\sigma(L)$ and the physical stability of the soliton is established by the following lemma (see \citep{chaos_nlde}):
\begin{lemma}\label{estabilidad_inestabilidad}
	Let $\sigma(L)$ be the spectrum of the operator defined in \eqref{sistema_inicial}. If there exists an eigenvalue $\Lambda \in \sigma(L)$ such that
	\begin{equation*}
		\text{Re}\left(\pm\sqrt{\beta_\pm \Lambda^2 + \rho^2}-\rho\right) > 0,
	\end{equation*}
	then the soliton is spectrally unstable.

	Conversely, if
	\begin{equation*}
		\text{Re}\left(\pm\sqrt{\beta_\pm \Lambda^2 + \rho^2}-\rho\right) \leq 0,
	\end{equation*}
	is satisfied for all $\Lambda \in \sigma(L)$, then the soliton is spectrally marginally stable.
\end{lemma}


The primary objective of this paper is to overcome the limitations of numerical and variational methods by providing a rigorous analytical proof that the subcritical
soliton $\Psi_-$ is unstable for all parameter values where it exists, i.e., across the entire subcritical domain $\lambda \in (0, \lambda_c)$. To overcome the indefiniteness of the Dirac operator, we propose a strategy that combines local asymptotic analyses at the boundaries of the parameter domain with global topological arguments.

Specifically, our approach is structured as follows: First, after characterizing the essential spectrum, we study the generalized null space of the operator
around the critical parameter $\lambda_c$. By restricting the problem to a parity-dependent subspace, we perform a local bifurcation analysis that reveals the onset of
instability through the splitting of the zero eigenvalue. Second, we analyze the non-relativistic limit ($\lambda \to 0$)
by squaring the spectral problem and reducing it, via a Lyapunov--Schmidt argument, to a P\"oschl--Teller Schr\"odinger operator. This yields a real pair of discrete eigenvalues and identifies the non-relativistic origin of the instability.
Finally, having established instability at both asymptotic boundaries, we use global topological continuation arguments to show that there exists an unstable eigenvalue that lies on the real
axis for all $\lambda \in (0, \lambda_c)$.

This paper is divided into the following sections:
\begin{enumerate}
	\item Analysis and determination of the essential spectra of the operator $L$.
	\item Calculus of the kernel and generalized null space of the operator $L$ and determination of the geometric and algebraic multiplicities of the zero eigenvalue.
	\item Restriction of the problem to a parity-dependent subspace of functions.
	\item First-order approximation of the splitting of the zero eigenvalue from the critical parameter, yielding a local instability result.
	\item Determination of the instability in the non-relativistic limit of $L$.
	\item Spectral estimates and global subcritical instability of the soliton for $(\lambda < \lambda_c)$, via topological arguments.
\end{enumerate}

\section{Essential spectra of $L$}

Before analyzing the discrete eigenvalues that determine stability, we must first map out the {essential spectrum} of the linearized operator $L$. This step is crucial for two reasons: it identifies the regions of the complex plane where wave packets disperse rather than grow, and it establishes the existence of a {spectral gap} near the origin where discrete instability modes can exist.

\begin{proposition}\label{espectro_essencial}
	Define the asymptotic parameter $\nu_1 = \sqrt{1-(1+\lambda^2)\Omega}$. Then, the essential spectrum of $L$ is given by
	\begin{equation}\label{essential_spectra}
		\sigma_{ess}(L) =
		\begin{cases}
			i(\mathbb{R}\setminus (-|\nu_1|, |\nu_1|)), & \nu_1^2 > 0, \\[2mm]
			i\mathbb{R},                                & \nu_1^2 = 0, \\[2mm]
			i\mathbb{R}\cup [-|\nu_1|,|\nu_1|],         & \nu_1^2 < 0.
		\end{cases}
	\end{equation}
\end{proposition}

\begin{proof}
	By Weyl's theorem (see e.g.\ Chapter XIII of \citet{reed_simon}), the essential spectrum of $L$ is invariant under relatively compact perturbations. This implies that it suffices to prove that the operator $L$ can be expressed as a compact perturbation of its asymptotic operator $L_\infty$, which is obtained by taking the limit of the coefficients as $|x| \to \infty$ (see \citep{Boussaid_Comech}).

	We decompose the operator as $L = L_\infty + R$, where
	\[
		L_\infty =
		\begin{pmatrix}
			0             & L_{0,\infty} \\
			-L_{1,\infty} & 0
		\end{pmatrix},
	\]
	and the asymptotic limits of the block operators are:
	\[
		L_{0,\infty} =
		\begin{pmatrix}
			\frac{1}{\lambda} - \frac{1+\lambda^2}{\lambda}\Omega & \partial_x                                   \\
			-\partial_x                                           & -\lambda - \frac{1+\lambda^2}{\lambda}\Omega
		\end{pmatrix},
		\qquad
		L_{1,\infty} =
		\begin{pmatrix}
			\lambda     & \partial_x         \\
			-\partial_x & -\frac{1}{\lambda}
		\end{pmatrix}.
	\]
	The remainder $R$ contains the spatially dependent functions $\psi(x)$, $\chi(x)$, and $\tilde{N}(x)$, all of which vanish as $|x| \to \infty$. Consequently, $R$ acts as a multiplication operator by a matrix-valued function with entries in $C^\infty(\mathbb{R})$.

	To prove that $R$ is $L_\infty$-compact, let $\theta_j$ be a smooth cutoff function such that $\theta_j(x) = 1$ for $|x| \le j\in \mathbb{N}$ and $\operatorname{supp}(\theta_j) \subset [-2j, 2j]$. The operator $\theta_j R$ has compact support and maps
	\[
		H^1(\mathbb{R}) \xrightarrow{\text{id}} H^1([-2j, 2j]) \hookrightarrow L^2([-2j, 2j]) \hookrightarrow L^2(\mathbb{R}).
	\]
	By the Rellich--Kondrachov theorem (Theorem 9.16 of \citet{brezis2011}), this embedding is compact. Furthermore, since the coefficients of $R$ decay at infinity, we have
	\[
		\|(1-\theta_j)R\|_{H^1 \mapsto L^2} \le \sup_{|x|>j} \|R(x)\|_\infty \xrightarrow{j\to\infty} 0.
	\]
	Thus, $R$ is the limit of a sequence of compact operators, implying $R$ is compact relative to $L_\infty$. By Weyl's theorem, we conclude $\sigma_{ess}(L) = \sigma_{ess}(L_\infty)$.

	We now calculate the spectrum of the constant-coefficient operator $L_\infty$ using the Fourier transform. The eigenvalue equation $(L_\infty - \Lambda I)f = 0$ becomes
	\begin{equation}\label{fourier}
		(A_\infty(\xi) - \Lambda I) \hat{f}(\xi) = 0,
	\end{equation}
	where $\hat{f}$ represents the Fourier transform of $f$ and the symbol $A_\infty(\xi)$ is defined by the matrix
	\[
		A_\infty(\xi) =
		\begin{pmatrix}
			0        & 0                 & \frac{1}{\lambda} - \frac{1+\lambda^2}{\lambda}\Omega & i\xi                                         \\[1mm]
			0        & 0                 & -i\xi                                                 & -\lambda - \frac{1+\lambda^2}{\lambda}\Omega \\[1mm]
			-\lambda & -i\xi             & 0                                                     & 0                                            \\[1mm]
			i\xi     & \frac{1}{\lambda} & 0                                                     & 0
		\end{pmatrix}.
	\]

	First, we show that the point spectrum is empty. If $\Lambda \in \sigma_p(L_\infty)$, there must exist a non-zero $\hat{f} \in L^2$ satisfying Eq.~\eqref{fourier}.
	However, this requires that $\det(A_\infty(\xi) - \Lambda I) = 0$ for all $\xi \in \mathbb{R}$. Since the eigenvalues of $A_\infty$ are non-constant functions of $\xi$,
	this gives a contradiction. Thus, no $L^2$--eigenfunction can be supported on this set, implying $\sigma_p(L_\infty) = \emptyset$.

	Similarly, the residual spectrum is empty because the operator possesses the symmetry $L_\infty^\dagger = L_\infty^T$ that precludes residual eigenvalues.

	Therefore, the spectrum is purely continuous and given by:
	\[
		\sigma(L_\infty) = \sigma_{ess}(L_\infty) = \bigcup_{\xi \in \mathbb{R}} \sigma(A_\infty(\xi)).
	\]
	Calculating the eigenvalues of the $4\times 4$ matrix $A_\infty(\xi)$ directly yields \eqref{essential_spectra}.
\end{proof}

\section{The kernel and the generalized null space of $L$}

Having established the location of the essential spectrum, we now turn our attention to the discrete spectrum. Of particular importance is the kernel (the null space) of $L$, as the zero eigenvalue often serves as the bifurcation point for instabilities.

\begin{lemma}
	Consider the operators $L$, $L_0$, and $L_1$ defined in \eqref{sistema_inicial}. The kernel of $L$ decomposes as:
	\begin{equation*}
		\ker(L) = \{(u, v)^T \in H^1(\mathbb{R})^4 \mid u \in \ker(L_1),\; v \in \ker(L_0)\}.
	\end{equation*}
\end{lemma}

\begin{proof}
	By definition, an element $y = (u, v)^T \in H^1(\mathbb{R})^4$ belongs to the kernel of $L$ if and only if $Ly = 0$. Using the block structure of $L$ defined in \eqref{sistema_inicial}, this condition reads:
	\begin{equation*}
		L y = \begin{pmatrix} 0 & L_0 \\ -L_1 & 0 \end{pmatrix} \begin{pmatrix} u \\ v \end{pmatrix} = \begin{pmatrix} L_0 v \\ -L_1 u \end{pmatrix} = \begin{pmatrix} 0 \\ 0 \end{pmatrix}.
	\end{equation*}
	This is equivalent to the decoupled system of equations $L_0 v = 0$ and $L_1 u = 0$. Therefore, $v \in \ker(L_0)$ and $u \in \ker(L_1)$.
\end{proof}

This lemma reduces the study of the zero mode of the full $4 \times 4$ system to the analysis of the kernels of the self-adjoint operators $L_0$ and $L_1$. We analyze these operators separately, as they correspond to distinct physical symmetries of the system.

\subsection{The operator $L_1$ and translational invariance}

The operator $L_1$ is associated with the spatial translational invariance of the NLDE. Its kernel is generated by the derivative of the soliton profile.

\begin{lemma}\label{kerL1}
	Let $\phi(x) = (\psi(x), \chi(x))^T$ be the soliton profile \eqref{soliton_profile}. Then the kernel of the self-adjoint operator $L_1$ is one-dimensional and is given by
	\begin{equation*}
		\ker(L_1) = \Span\{\phi_x\},
	\end{equation*}
	Moreover, its essential spectrum is
	\begin{equation*}
		\sigma_{ess}(L_1) = \left(-\infty, -1/\lambda\right] \cup \left[\lambda, +\infty\right).
	\end{equation*}
\end{lemma}

\begin{proof}
	Direct calculation confirms that $L_1 \phi_x = 0$ for all $\lambda \in (0,1)$, which reflects the translational invariance of the system.

	We now establish that the kernel contains at most one element (up to scalar multiplication). This follows from the asymptotic behavior imposed to the eigenfunctions associated with the zero eigenvalue. In this setting, any eigenfunction must satisfy the decay condition $\varphi(x) \sim e^{-\mu x}$ as $x \to \infty$, for a specific value $\mu > 0$ depending on $\lambda$ and $\Omega$.

	Suppose there exists another linearly independent eigenfunction $\bar{\varphi} \sim e^{-\mu x}$. Under this hypothesis, the linear combination $\varphi - \bar{\varphi}$
	would belong to the kernel but would fail to exhibit the required exponential decay, leading to a contradiction (see e.g. \citep{Berkolaiko_comech}). Thus, the kernel is one-dimensional.

	The characterization of $\sigma_{ess}(L_1)$ follows from the same compact-perturbation and Fourier-symbol argument used in Proposition~\ref{espectro_essencial}, applied to the asymptotic operator
	\[
		L_{1,\infty} =
		\begin{pmatrix}
			\lambda     & \partial_x \\
			-\partial_x & -1/\lambda
		\end{pmatrix}.
	\]
\end{proof}

\subsection{The operator $L_0$ and phase invariance}

The spectral properties of $L_0$ are more intricate. This operator is associated with the $U(1)$ phase symmetry of the problem, and its spectral analysis requires a more delicate treatment. We derive its full spectrum by explicitly solving the associated differential equations.

\begin{proposition}\label{spectrum_L0}
	The spectrum of $L_0$ consists of an essential spectrum
	\begin{equation*}
		\sigma_{ess}(L_0) = \left(-\infty, -\lambda+\mu_0\right] \cup \left[1/\lambda+\mu_0, +\infty\right),
	\end{equation*}
	and exactly two simple discrete eigenvalues
	\begin{equation*}
		\sigma_d(L_0) =\{\mu_1, \mu_0\} ,
	\end{equation*}
	where $\mu_0 = -\left(\lambda - \frac{1}{\lambda} + \frac{1+\lambda^2}{\lambda}\Omega\right)$ and $\mu_1 = -\frac{1+\lambda^2}{\lambda}\Omega$.

	Furthermore, the eigenfunctions associated with the discrete spectrum are
	\begin{equation}\label{eigenfunctions_L0}
		\mu_0 \in \sigma_p(L_0) \implies
		\begin{pmatrix}
			\psi \\ \chi
		\end{pmatrix},
		\qquad
		\mu_1 \in \sigma_p(L_0) \implies
		\begin{pmatrix}
			\chi \\ \psi
		\end{pmatrix}.
	\end{equation}

\end{proposition}

\begin{proof}
	We consider the eigenvalue problem
	\begin{equation}\label{problemaL0}
		L_0
		\begin{pmatrix}
			f_\Lambda \\ g_\Lambda
		\end{pmatrix}
		= \Lambda
		\begin{pmatrix}
			f_\Lambda \\ g_\Lambda
		\end{pmatrix},
	\end{equation}
	which can be written explicitly as the system:
	\begin{equation*}
		\begin{split}
			\partial_x f_\Lambda & =
			\left(\tilde{N}(x) - \lambda - \frac{1+\lambda^2}{\lambda}\Omega - \Lambda\right) g_\Lambda, \qquad
			\partial_x g_\Lambda =
			\left(\tilde{N}(x) - \frac{1}{\lambda} + \frac{1+\lambda^2}{\lambda}\Omega + \Lambda\right) f_\Lambda .
		\end{split}
	\end{equation*}

	Differentiating the system with respect to $x$ allows us to uncouple the equations. We obtain:
	\begin{equation*}
		\begin{split}
			\partial_x^2 f_\Lambda & =
			\left(\tilde{N}(x) - \lambda - \frac{1+\lambda^2}{\lambda}\Omega - \Lambda\right)
			\left(\tilde{N}(x) - \frac{1}{\lambda} + \frac{1+\lambda^2}{\lambda}\Omega + \Lambda\right) f_\Lambda
			+ \tilde{N}'(x) g_\Lambda, \\
			\partial_x^2 g_\Lambda & =
			\left(\tilde{N}(x) - \lambda - \frac{1+\lambda^2}{\lambda}\Omega - \Lambda\right)
			\left(\tilde{N}(x) - \frac{1}{\lambda} + \frac{1+\lambda^2}{\lambda}\Omega + \Lambda\right) g_\Lambda
			+ \tilde{N}'(x) f_\Lambda.
		\end{split}
	\end{equation*}
	This structure implies that the linear combinations $(f_\Lambda - g_\Lambda)$ and $(f_\Lambda + g_\Lambda)$ satisfy the following decoupled differential equations:
	\begin{equation}
		\begin{split}
			\partial_x^2 (f_\Lambda - g_\Lambda) & =
			\left[ \left(\tilde{N}(x) - \lambda - \frac{1+\lambda^2}{\lambda}\Omega - \Lambda\right)
				\left(\tilde{N}(x) - \frac{1}{\lambda} + \frac{1+\lambda^2}{\lambda}\Omega + \Lambda\right)
			- \tilde{N}'(x) \right] (f_\Lambda - g_\Lambda), \\
			\partial_x^2 (f_\Lambda + g_\Lambda) & =
			\left[ \left(\tilde{N}(x) - \lambda - \frac{1+\lambda^2}{\lambda}\Omega - \Lambda\right)
				\left(\tilde{N}(x) - \frac{1}{\lambda} + \frac{1+\lambda^2}{\lambda}\Omega + \Lambda\right)
				+ \tilde{N}'(x) \right] (f_\Lambda + g_\Lambda).
		\end{split}
	\end{equation}

	Observe that the second equation is related to the first via the reflection $x \mapsto -x$ (since $\tilde{N}(x)$ is even and $\tilde{N}'(x)$ is odd). Therefore, it suffices to determine the general solution of the first differential equation, which is
	\begin{equation*}
		\begin{split}
			f_\Lambda(x) - g_\Lambda(x) & = a_1 \Upsilon_+(x) + a_2 \Upsilon_-(x),   \\
			f_\Lambda(x) + g_\Lambda(x) & = b_1 \Upsilon_+(-x) + b_2 \Upsilon_-(-x),
		\end{split}
	\end{equation*}
	where $a_1, a_2, b_1, b_2 \in \mathbb{C}$ and $\Upsilon_\pm$ are defined as
	\begin{equation*}
		\Upsilon_\pm (x) = \sqrt{(a-1)+\left(\frac{\lambda \sech(x)}{1+\lambda \tanh(x)}\right)^2}
		\exp\left\{\pm\left(bx-\atanh\left[\frac{a(1+\lambda \tanh(x))-1}{b \lambda}\right]\right)\right\},
	\end{equation*}
	with parameters
	\begin{equation*}
		\begin{split}
			a & = \frac{1}{1-\lambda^2}\left((1+\lambda^2)\Omega+\lambda \Lambda\right)
			\left(-(1-\lambda^2)+(1+\lambda^2)\Omega+\lambda\Lambda\right)-1,            \\
			b & = \frac{1}{\lambda}\sqrt{(\lambda^2+(1+\lambda^2)\Omega+\lambda \Lambda)
				(1-(1+\lambda^2)\Omega-\lambda \Lambda)}.
		\end{split}
	\end{equation*}

	For $\Lambda$ to belong to the discrete spectrum, we require the eigenfunctions to satisfy $(f_\Lambda, g_\Lambda)^T \in H^1(\mathbb{R})^2$. Based on the asymptotic behavior of $\Upsilon_\pm$, this condition requires $a = 1$. This constraint is satisfied if and only if $\Lambda = \mu_0$ or $\Lambda = \mu_1$. In these cases, the solutions become \eqref{eigenfunctions_L0}. This proves that the discrete spectrum is $\sigma_d(L_0) = \{\mu_0, \mu_1\}$.

	The essential spectrum (which coincides with the continuous spectrum) follows from an argument analogous to the one used in Proposition~\ref{espectro_essencial}, so we omit it here.
\end{proof}

The explicit form of the eigenvalues allows us to track the behavior of the kernel as a function of the parameter $\lambda$, so we have the following corollary:
\begin{corollary}\label{kerL0}
	The dimension of the kernel of $L_0$ depends on the parameter $\lambda$:
	\begin{itemize}
		\item If $\lambda = \lambda_c$, then $\mu_0 = 0$ and $\ker(L_0) = \Span\{\phi\}$.
		\item If $\lambda \neq \lambda_c$, then there exists an open ball $B_R(\lambda)$ centered at zero such that $B_R(\lambda) \cap \sigma(L_0) = \emptyset$.
	\end{itemize}
\end{corollary}

This result is fundamental for the stability analysis. It implies that for $\lambda \neq \lambda_c$, the geometric multiplicity of the zero eigenvalue of the full operator $L$ is exactly 1 (coming solely from $L_1$). However, at the critical point $\lambda_c$, the multiplicity jumps to 2.

\subsection{Generalized null space away from criticality ($\lambda \neq \lambda_c$)}

While the kernel describes the geometric multiplicity of the zero eigenvalue, the stability is determined by its {algebraic} multiplicity. If the algebraic multiplicity exceeds the geometric multiplicity, the operator possesses a generalized null space $N_g$, consisting of Jordan chains. The existence of these chains is often the precursor to oscillatory instabilities, \citep{spectral_dyinamical_stability_kapitula, grillakis1987}.

The generalized null space is defined by
\begin{equation*}
	N_g = \bigcup_{k=1}^\infty \ker(L^k), \qquad L^k = \underbrace{L \circ \cdots \circ L}_{k}.
\end{equation*}

For the non-critical case, the kernel is generated solely by the translational symmetry. The following proposition characterizes the Jordan chain associated with this mode:

\begin{proposition}\label{null_space_noncritical}
	Assume $\lambda \neq \lambda_c$ and consider $\tilde{\phi}_x \coloneqq -L_0^{-1} \phi_x$. The structure of the generalized null space depends on the orthogonality of the translational mode:
	\begin{enumerate}
		\item If $\langle \phi_x, \tilde{\phi}_x \rangle \neq 0$, the generalized null space is two-dimensional
		      \begin{equation*}
			      N_g = \Span\{\Psi_1, \Psi_2\}, \quad \dim(N_g) = 2.
		      \end{equation*}
		      The basis vectors are $\Psi_1 = (\phi_x, \vec{0})^T$ (the kernel) and $\Psi_2 = (\vec{0}, \tilde{\phi}_x)^T$ (the generalized eigenvector).

		\item If $\langle \phi_x, \tilde{\phi}_x \rangle = 0$, the chain extends further, and
		      \begin{equation*}
			      \Span\{\Psi_1, \Psi_2\} \subset N_g,\qquad	\dim(N_g) \geq 3.
		      \end{equation*}
	\end{enumerate}
\end{proposition}

\begin{proof}
	We will construct the Jordan chain iteratively. By Lemma~\ref{kerL1}, we already know that $\ker(L) = \Span\{\Psi_1\}$.

	To find the first generalized eigenvector, we look for $\Psi \in \ker(L^2)$ such that $L\Psi = -\Psi_1 \neq 0$. Writing $\Psi = (\varphi, \tilde{\varphi})^T$, the equation $L^2 \Psi = 0$ is equivalent to the system:
	\begin{equation*}
		L_0 L_1 \varphi = 0, \qquad L_1 L_0 \tilde{\varphi} = 0.
	\end{equation*}

	From Corollary \ref{kerL0}, since $\lambda \neq \lambda_c$, $L_0$ is invertible. Thus, $L_0(L_1 \varphi) = 0$ implies $L_1 \varphi = 0$, so $\varphi \in \ker(L_1) = \Span\{\phi_x\}$.
	For the second component, $L_1 L_0 \tilde{\varphi} = 0$ implies $L_0 \tilde{\varphi} \in \ker(L_1)$. Since $L_0$ is invertible, we find the unique solution $\tilde{\varphi} = -L_0^{-1} \phi_x \coloneqq \tilde{\phi}_x$. Thus, $\Psi_2 = (\vec{0}, \tilde{\phi}_x)^T$ exists.

	Next, we look for a vector $\Psi_3 \in \ker(L^3)$	
such that $L \Psi_3 = -\Psi_2$. This leads to the equation $L_1 \tilde{\varphi}_2 = \tilde{\phi}_x$. By the Fredholm alternative \citep{brezis2011}, since $L_1$ is self-adjoint, this equation is solvable if and only if the source term is orthogonal to the kernel:
	\begin{equation*}
		\langle \phi_x, \tilde{\phi}_x \rangle = 0.
	\end{equation*}
	If the last inner product is non-zero, the chain terminates at length 2.
\end{proof}


\subsection{The critical generalized null space ($\lambda = \lambda_c$)}

According to Corollary~\ref{kerL0}, at the critical parameter $\lambda_c$, the kernel of $L_0$ becomes non-trivial ($\ker(L_0) = \text{span}\{\phi\}$).
Consequently, the geometric multiplicity of the zero eigenvalue of $L$ is equal to 2 and we must determine if new Jordan chains emerge.

\begin{proposition}\label{null_space_critical}
	At the bifurcation point $\lambda = \lambda_c$, the generalized null space is the direct sum of two Jordan chains:
	\begin{enumerate}
		\item If $\langle \phi_x, \tilde{\phi}_x \rangle \neq 0$, then
		      \begin{equation*}
			      N_g = \Span\{\Phi_1, \Phi_2\} \oplus \Span\{\Psi_1, \Psi_2\}, \quad \dim(N_g) = 4,
		      \end{equation*}
		      where the phase chain basis vectors are
		      \begin{equation*}
			      \Phi_1 = (\vec{0}, \phi)^T, \qquad \Phi_2 = (\tilde{\phi}, \vec{0})^T,
		      \end{equation*}
		      and the translational chain basis vectors are
		      \begin{equation*}
			      \Psi_1 = ({\phi}_x, \vec{0})^T, \qquad \Psi_2 = (\vec{0}, \tilde{\phi}_x)^T.
		      \end{equation*}
		      The generalized eigenvector $\tilde{\phi}$ is explicitly given by:
		      \begin{equation}\label{tilde_phi}
			      \begin{split}
				      \tilde{\phi} & \coloneqq -L_1^{-1} \phi
				      = \frac{1+\lambda^2}{2}\partial_\lambda \phi
				      + \frac{1-\lambda^2}{2\lambda}
				      \left. \partial_\varepsilon \left(\sqrt{\varepsilon} \phi(\varepsilon x)\right)\right|_{\varepsilon=1} \\[4pt]
				                   & =
				      \frac{\left( 1 + \lambda^2 \right)\,\sech(x)}
				      {2\sqrt{2\lambda}\left(1 - \lambda^2 + \lambda^2\sech^2(x)\right)^2}
				      \begin{pmatrix}
					      1+ 3\lambda^2 \tanh^2(x)
					      \\[8pt]
					      {\lambda}\,\left(3 + \lambda^2 \tanh^2(x)\right)\tanh(x)
				      \end{pmatrix}.
			      \end{split}
		      \end{equation}
		\item If $\langle \phi_x, \tilde{\phi}_x \rangle = 0$, the dimension increases further:
		      \begin{equation*}
			      \Span\{\Phi_1, \Phi_2\} \oplus \Span\{\Psi_1, \Psi_2\} \subset N_g, \qquad \dim(N_g) \geq 5.
		      \end{equation*}
	\end{enumerate}
\end{proposition}

\begin{proof}
	The proof of this proposition will resemble that of Proposition \ref{null_space_noncritical}. For this purpose, it is interesting to notice that at $\lambda = \lambda_c$, $\ker(L) = \text{span}\{\Phi_1, \Psi_1\}$. We will examine the extension of these kernels into Jordan chains:

	\medskip
	\noindent\textit{The translational chain.}
	In this case, we obtain the same result as in Proposition \ref{null_space_noncritical}, due to the fact that the Jordan chain generated by $\Psi_1$ is always orthogonal to $\Phi_1$ because of parity restrictions.

	\medskip
	\noindent\textit{The phase chain.}
	We look for $\Phi_2$ such that $L \Phi_2 = \Phi_1 = (\vec{0}, \phi)^T$. This requires solving:
	\begin{equation*}
		\begin{pmatrix} 0 & L_0 \\ -L_1 & 0 \end{pmatrix} \begin{pmatrix} u \\ v \end{pmatrix} = \begin{pmatrix} 0 \\ \phi \end{pmatrix}.
	\end{equation*}
	The first row $L_0 v = 0$ is satisfied by $v=0$. The second row becomes $L_1 u = -\phi$. Since $\phi$ is orthogonal to $\ker(L_1) = \text{span}\{\phi_x\}$ (parity argument), a unique solution $\tilde{\phi} = -L_1^{-1}\phi$ exists. The explicit expression in Eq.~\eqref{tilde_phi} follows from the differentiation of Eq.~\eqref{sistema_psi_chi} with respect to $\lambda$ and the scaling factor.

	To check if this chain extends to a third vector ($\ker(L^3)$), we would need to solve $L \Psi = \Phi_2$, which implies $L_0 v = \tilde{\phi}$. This requires $\tilde{\phi} \perp \ker(L_0)$, i.e., $\langle \phi, \tilde{\phi} \rangle = 0$.

	We calculate this inner product explicitly:
	\begin{equation*}
		\langle \phi, \tilde{\phi} \rangle
		=
		\left\langle
		\phi,
		\frac{1+\lambda^2}{2}\partial_\lambda \phi
		+ \frac{1-\lambda^2}{2\lambda}
		\left. \partial_\varepsilon \left(\sqrt{\varepsilon} \phi(\varepsilon x)\right)\right|_{\varepsilon=1}
		\right\rangle
		= \frac{1+\lambda^2}{4}\partial_\lambda \|\phi\|_2^2 = \frac{(1+\lambda_c^2)^2}{(1-\lambda_c^2)^2}\neq 0.
	\end{equation*}
	Thus, the orthogonality condition fails, and the chain terminates at length 2.
\end{proof}

\section{Parity-defined spinor subspace}

A significant challenge in analyzing the stability of solitons is handling the zero eigenvalue associated with translational invariance. This mode is present for all values of $\lambda$, as stated in Propositions \ref{null_space_noncritical} and \ref{null_space_critical}, but it does not contribute to the type of instability we are investigating. Therefore, to focus on the spectral splitting that destabilizes the subcritical $\Psi_-$ soliton, we must filter out this translational mode from our analysis.

Fortunately, the linearized system \eqref{sistema_inicial} possesses underlying parity symmetries that naturally decompose the functional space into invariant subspaces. By restricting the eigenvalue problem to a specific parity-dependent subspace, we can isolate the eigenvalues responsible for the instability.

\begin{definition}
	Let $a_i : \mathbb{R} \mapsto \mathbb{C}$ for $i\in \{1, ..., 2N\}$. A vector $\varphi = (a_1, a_2, \dots, a_{2N})^T$ is said to be \textit{positively alternating} $(p.a.)$ if the components $a_{2i-1}$ are even and $a_{2i}$ are odd functions for $i = 1,\dots,N$. Conversely, it is \textit{negatively alternating} $(n.a.)$ if $a_{2i-1}$ are odd and $a_{2i}$ are even functions.
\end{definition}

This definition allows us to decompose the total Hilbert space based on the spatial symmetry of its components.

\begin{lemma}
	The Hilbert space $H^1(\mathbb{R})^4$ admits the following direct sum decomposition:
	\begin{equation*}
		H^1(\mathbb{R})^4 = H_{p.a.} \oplus H_{n.a.},
	\end{equation*}
	where $H_{p.a.}$ and $H_{n.a.}$ denote the subspaces of elements satisfying the corresponding alternating symmetry.
\end{lemma}

We formalize this decomposition through the introduction of a parity operator.

\begin{definition}
	The alternating parity operator $P : H^1(\mathbb{R})^4 \mapsto H^1(\mathbb{R})^4$ is defined by its action on the decomposition
	\begin{equation*}
		P(\varphi_{p.a.} + \varphi_{n.a.}) = \varphi_{p.a.} - \varphi_{n.a.}.
	\end{equation*}
\end{definition}

This definition naturally induces the orthogonal projectors
\begin{equation*}
	P_+ = \frac{1}{2}(I + P), \qquad P_- = \frac{1}{2}(I - P),
\end{equation*}
where $P_+ : H^1(\mathbb{R})^4 \mapsto H_{p.a.}$ and $P_- : H^1(\mathbb{R})^4 \mapsto H_{n.a.}$.

\begin{proposition}
	The linearized operator $L$ commutes with the parity operator $P$; that is, $[L,P] = 0$. Consequently, $L$ leaves the subspaces invariant and admits the block diagonal form
	\begin{equation*}
		L = P_+ L P_+ + P_- L P_- = L_p + L_n,
	\end{equation*}
	where $L_p = L|_{H_{p.a.}}$ and $L_n = L|_{H_{n.a.}}$ are the restrictions of $L$ to each subspace.
\end{proposition}

This reduction is structurally crucial. The translational modes $\phi_x$ (and its generalized partners) inherently possess negatively alternating parity ($n.a.$). Therefore, the translational kernel is confined entirely within $L_n$. By restricting our analysis to $L_p$, we effectively \textit{quotient out} the translational symmetry and focus solely on the instability splitting.

Thus, Propositions \ref{null_space_noncritical} and \ref{null_space_critical} lead to the following result:
\begin{theorem}
	Consider the restricted operator $L_p : H_{p.a.} \mapsto H_{p.a.}$. Its spectral properties near zero characterize the splitting:
	\begin{itemize}
		\item If $\lambda \neq \lambda_c$: The kernel is trivial, i.e.,
		      \begin{equation*}
			      \ker(L_p) = N_g(L_p) = \{0\}.
		      \end{equation*}
		      Consequently, the geometric ($g.m.$) and algebraic ($a.m.$) multiplicities are both 0.

		\item If $\lambda = \lambda_c$: The kernel and generalized null space become non-trivial due to the resonance of the phase mode
		      \begin{equation*}
			      \ker(L_p) = \Span\{\Phi_1\}, \quad N_g(L_p) = \Span\{\Phi_1, \Phi_2\}.
		      \end{equation*}
		      This corresponds to multiplicities: $g.m. = 1$ and $a.m. = 2$.
		      The basis vectors are explicitly given by:
		      \begin{equation*}
			      \Phi_1 = (\vec{0}, \phi)^T, \qquad
			      \Phi_2 = (\tilde{\phi}, \vec{0})^T,
		      \end{equation*}
		      where $\tilde{\phi}$ is defined as
		      \begin{equation*}
			      \tilde{\phi} = -L_1^{-1} \phi
			      = \frac{1+\lambda^2}{2} \partial_\lambda \phi
			      + \frac{1-\lambda^2}{2\lambda}
			      \left. \partial_\varepsilon \bigl(\sqrt{\varepsilon} \phi(\varepsilon x)\bigr) \right|_{\varepsilon=1}.
		      \end{equation*}
	\end{itemize}
\end{theorem}


As a consequence of this theorem, the zero eigenvalue of the restricted operator $L_p$ undergoes a bifurcation at $\lambda = \lambda_c$. The jump in algebraic multiplicity from 0 to 2 indicates the collapse of two eigenvalues at the origin. This structure implies that, as $\lambda$ deviates from $\lambda_c$, these two eigenvalues emerge from zero, becoming either a pair of purely real (unstable) or purely imaginary (stable) eigenvalues, constrained by the symmetry of the spectrum.

\section{Spectral splitting near the critical parameter}

Having characterized the algebraic structure of the null space at $\lambda_c$, we now turn to the question: how do these zero modes evolve when the parameter $\lambda$ is changed? To answer this, we must track the eigenvalues as they move away from the origin.

First, we must ensure that the eigenvalues of interest are not embedded in the continuous spectrum. The following lemma guarantees the existence of a region around the origin, free from the interference of scattering modes.

\begin{lemma}
	\label{hipotesis}
	For $\lambda \in (0, \lambda_c)$, there exists $R(\lambda) > 0$ such that
	\[
		\dim(B_R \cap \sigma(L_p)) < \infty,
	\]
	where $B_R$ is the ball centered at zero and of radius $R(\lambda)$.
	In other words, the eigenvalues in the vicinity of zero are discrete and isolated.
\end{lemma}

The validity of this lemma follows directly from Proposition \ref{espectro_essencial}. Since the spectrum is the union of the discrete and essential
spectra ($\sigma = \sigma_d \cup \sigma_{ess}$), then $0\in \sigma_d(L)$ and thus, there exists a sufficiently small
ball $B_R$ that contains only a finite number of discrete eigenvalues counting multiplicity.

For the local analysis near $\lambda_c$, we use the algebraic description of the critical Jordan chain together with a Lyapunov--Schmidt reduction for Fredholm operators
(see e.g. Chapter VIII \citet{kato}). By separating the problem into the critical directions and their complement, we can explicitly obtain the trajectory of the eigenvalues as a
function of the perturbation $\lambda - \lambda_c$.

\begin{theorem}\label{teorema_bifurcacion}
	For $\delta > 0$ sufficiently small and $\lambda \in (\lambda_c-\delta, \lambda_c+\delta)$, the operator $L_p$ possesses two eigenvalues $\Lambda(\lambda)$ that emerge from the origin. Their asymptotic behavior is given by:
	\begin{equation*}
		\Lambda
		=  \pm i\sqrt{\frac{2(1+\Omega)\|\phi\|_2^2}{|\langle \phi \mid \tilde{\phi} \rangle|}}\,\sqrt{\lambda - \lambda_c} + O(|\lambda-\lambda_c|).
	\end{equation*}
\end{theorem}

\begin{proof}
	The proof relies on a finite-dimensional Lyapunov--Schmidt reduction around the critical point. Since $L_p$ is not self-adjoint, we must use the $L^2$-adjoint operator and establish the biorthogonality relations of the critical modes.

	The $L^2$-adjoint operator restricted to the positively alternating subspace is
	\begin{equation}\label{eq_adjoint_Lp}
		L_p^\dagger = \begin{pmatrix} 0 & -L_{1,p} \\ L_{0,p} & 0 \end{pmatrix}.
	\end{equation}
	At $\lambda = \lambda_c$, the generalized null space of $L_p^\dagger$ is spanned by the adjoint vectors:
	\begin{equation}\label{eq_adjoint_eigenvectors}
		\Phi_1^\dagger = (\phi, \vec{0})^T, \qquad \Phi_2^\dagger = (\vec{0}, \tilde{\phi})^T.
	\end{equation}
	Direct calculation from \eqref{eq_adjoint_Lp} confirms that $L_p^\dagger \Phi_1^\dagger = 0$ and $L_p^\dagger \Phi_2^\dagger = \Phi_1^\dagger$. Together with the basis vectors $\Phi_1, \Phi_2$ of the generalized null space of $L_p$ from Proposition~\ref{null_space_critical}, we have the following biorthogonality relations:
	\begin{equation}\label{biorthogonality}
		\langle \Phi_1^\dagger \mid \Phi_1 \rangle = 0, \quad
		\langle \Phi_2^\dagger \mid \Phi_2 \rangle = 0, \quad
		\langle \Phi_1^\dagger \mid \Phi_2 \rangle = \langle \Phi_2^\dagger \mid \Phi_1 \rangle = \langle \phi \mid \tilde{\phi} \rangle \neq 0.
	\end{equation}

	We seek nontrivial solutions to the eigenvalue equation near the critical point:
	\begin{equation}\label{eq_eigenvalue_problem}
		(L_p(\lambda) + \Lambda)\psi = 0.
	\end{equation}
	We decompose the Hilbert space as $H_{p.a.} = \ker(L_p(\lambda_c)) \oplus X_1$, where the complement is defined as $X_1 \coloneqq \{u \in H_{p.a.} : \langle \Phi_2^\dagger \mid u \rangle = 0\}$. By \eqref{biorthogonality}, $\langle \Phi_2^\dagger \mid \Phi_1 \rangle \neq 0$, ensuring this is a direct sum. We normalize the eigenvector as $\psi = \Phi_1 + w$ with $w \in X_1$.

	Let $P$ denote the projection onto $\operatorname{Ran}(L_p(\lambda_c)) = \{v \in H_{p.a.} : \langle \Phi_1^\dagger \mid v \rangle = 0\}$. Projecting \eqref{eq_eigenvalue_problem} onto this range and its complement yields the coupled system:
	\begin{align}
		P(L_p(\lambda) + \Lambda)(\Phi_1 + w)                                                                  & = 0, \label{eq_range}       \\
		g(\lambda, \Lambda) \coloneqq \langle \Phi_1^\dagger \mid (L_p(\lambda) + \Lambda)(\Phi_1 + w) \rangle & = 0. \label{eq_bifurcation}
	\end{align}
	At $(\lambda, \Lambda, w) = (\lambda_c, 0, 0)$, the linearization of \eqref{eq_range} with respect to $w$ is exactly the restriction $PL_p(\lambda_c)|_{X_1}$. Since $L_p$ is a Fredholm operator of index zero and its kernel intersects $X_1$ trivially, this map is a bijection. By the implicit function theorem, there exists a unique smooth map $w_\ast(\lambda, \Lambda)$ satisfying \eqref{eq_range} locally, with $w_\ast(\lambda_c, 0) = 0$.

	Substituting $w = w_\ast(\lambda, \Lambda)$ into \eqref{eq_bifurcation} reduces the problem to the scalar bifurcation equation $g(\lambda, \Lambda) = 0$. We evaluate its derivatives at the critical point $(\lambda_c, 0)$, keeping the basis vectors fixed.

	Since $\Phi_1 \in \ker(L_p(\lambda_c))$, it immediately follows that $g(\lambda_c, 0) = 0$. For the first-order derivatives, we use $L_p^\dagger \Phi_1^\dagger = 0$ and \eqref{biorthogonality} to obtain:
	\begin{equation}\label{eq_dg_dLambda}
		\partial_\Lambda g\big|_{(\lambda_c,0)} = \langle \Phi_1^\dagger \mid \Phi_1 \rangle + \langle L_p^\dagger \Phi_1^\dagger \mid \partial_\Lambda w_\ast \rangle = 0.
	\end{equation}
	For the parameter derivative, the explicit form of $\partial_\lambda L_p$ yields $(\partial_\lambda L_p)\Phi_1 = ((\partial_\lambda L_{0,p})\phi,\, \vec{0})^T$. Thus:
	\begin{equation}\label{eq_dg_dlambda}
		\partial_\lambda g\big|_{(\lambda_c,0)} = \langle \Phi_1^\dagger \mid \left.(\partial_\lambda L_p)\right|_{\lambda_c}\, \Phi_1 \rangle = \langle \phi \mid \left.(\partial_\lambda L_{0,p})\right|_{\lambda_c} \phi \rangle.
	\end{equation}
	Using the eigenvalue relation $L_{0,p}(\lambda)\phi(\lambda) = \mu_0(\lambda)\phi(\lambda)$ and the fact that $\mu_0(\lambda_c) = 0$, we differentiate at $\lambda_c$ and project onto $\phi$ to find:
	\begin{equation}\label{eq_derivada_lambda}
		\partial_\lambda g\big|_{(\lambda_c,0)} = \mu_0'(\lambda_c)\|\phi\|_2^2,
	\end{equation}
	where a direct computation gives $\mu_0'(\lambda_c) = -2(1+\Omega) < 0$.

	To find the first non-vanishing term for $\Lambda$, we need the second derivative. Differentiating \eqref{eq_range} with respect to $\Lambda$ at $(\lambda_c,0)$ yields $P\Phi_1 + PL_p(\lambda_c) \partial_\Lambda w_\ast = 0$. From the Jordan chain relations, $L_p(\lambda_c)\Phi_2 = \Phi_1$, so $P\Phi_1 = P L_p(\lambda_c)\Phi_2$. Because $\partial_\Lambda w_\ast \in X_1$, we deduce $\partial_\Lambda w_\ast\big|_{(\lambda_c,0)} = -\Phi_2$. Substituting this into the second derivative of $g$ provides:
	\begin{equation}\label{eq_segunda_derivada}
		\partial_\Lambda^2 g\big|_{(\lambda_c,0)} = 2\langle \Phi_1^\dagger \mid \partial_\Lambda w_\ast \rangle = -2\langle \Phi_1^\dagger \mid \Phi_2 \rangle = -2\langle \phi \mid \tilde{\phi} \rangle \neq 0.
	\end{equation}

	Because $\partial_\lambda g \neq 0$, the implicit function theorem allows us to locally invert $g(\lambda(\Lambda), \Lambda) = 0$ as a smooth curve $\lambda(\Lambda)$ with $\lambda(0) = \lambda_c$. Differentiating the identity $g(\lambda(\Lambda), \Lambda) = 0$ implicitly with respect to $\Lambda$ yields $\lambda'(0) = 0$, and the second derivative is:
	\begin{equation}\label{eq_lambda_expansion}
		\lambda''(0) = -\frac{\partial_\Lambda^2 g}{\partial_\lambda g}\bigg|_{(\lambda_c,0)} = \frac{2\langle \phi \mid \tilde{\phi} \rangle}{\mu_0'(\lambda_c)\|\phi\|_2^2}.
	\end{equation}
	Thus, the parameter expands as $\lambda - \lambda_c = \frac{1}{2}\lambda''(0)\Lambda^2 + O(\Lambda^3)$. Inverting this expansion gives the eigenvalue splitting:
	\begin{equation}\label{eq_autovalores_splitting}
		\Lambda = \pm\sqrt{C(\lambda_c)\,(\lambda - \lambda_c)} + O(|\lambda-\lambda_c|), \qquad C(\lambda_c) = \frac{\mu_0'(\lambda_c)\|\phi\|_2^2}{\langle \phi \mid \tilde{\phi} \rangle}.
	\end{equation}
	Finally, we determine the sign of $C(\lambda_c)$. We already know $\mu_0'(\lambda_c) < 0$. For the denominator, using the explicit formula for $\tilde{\phi}$ from Proposition~\ref{null_space_critical}, we have:
	\begin{equation}\label{eq_inner_product_sign}
		\langle \phi \mid \tilde{\phi} \rangle = \frac{1+\lambda_c^2}{4}\left.\partial_\lambda \|\phi\|_2^2 \right|_{\lambda_c} = \frac{(1+\lambda_c^2)^2}{(1-\lambda_c^2)^2} > 0.
	\end{equation}
	Consequently, $C(\lambda_c) < 0$, and the eigenvalues take the purely imaginary form for $\lambda > \lambda_c$ and real form for $\lambda < \lambda_c$:
	\begin{equation*}
		\Lambda =  \pm i\sqrt{|C(\lambda_c)|}\,\sqrt{\lambda - \lambda_c} + O(|\lambda-\lambda_c|),
	\end{equation*}
	where $C(\lambda_c)$ is given in \eqref{eq_autovalores_splitting}.    \qedhere
\end{proof}

The nature of the eigenvalues near zero is determined by the term $\sqrt{\lambda - \lambda_c}$, which provides a local stability criterion.

This result has a crucial physical consequence: for $\lambda > \lambda_c$ (in a close neighborhood),
the bifurcated eigenvalues are purely imaginary, which do not produce instability.
However, for the subcritical regime $\lambda < \lambda_c$ (where $\lambda - \lambda_c < 0$), the factor $\sqrt{\lambda - \lambda_c}$
becomes imaginary, producing purely real eigenvalues. This establishes the instability of the system in the subcritical region $(\lambda_c-\delta, \lambda_c)$ for sufficiently small $\delta>0$, as previously predicted numerically by \citep{chaos_nlde}.

\section{Spectral enclosure and estimates}

While the previous sections focused on the local splitting of eigenvalues near the origin, establishing global stability requires control over the entire spectrum.
To this end, we derive an explicit bound for the spectrum of $L$. This result confines the eigenvalues to a vertical strip in the complex plane, providing an \textit{a priori} bound on the maximum possible growth rate of any instability.

\begin{proposition}\label{prop:banda}
	The spectrum $\sigma(L)$ is contained within the following vertical band:
	\begin{equation}\label{eq:banda}
		\sigma(L) \subset \{z \in \mathbb{C} : |\operatorname{Re}(z)| \leq W\},
	\end{equation}
	where
	\begin{equation*}
		W = |\mu_0| + 4\lambda \frac{1+\lambda^2}{(1-\lambda^2)^2},
	\end{equation*}
	and $\mu_0$ given in Proposition \ref{spectrum_L0}.
\end{proposition}

\begin{proof}
	The strategy is to transform the non-self-adjoint eigenvalue problem into a perturbation of a self-adjoint operator, allowing us to apply the resolvent estimates from \citet{Cuenin_Tretter}.

	First, we notice that the diagonal blocks $L_0$ and $L_1$ defined in \ref{def_operadores} are related by the following direct calculation:
	\begin{equation*}
		L_1 = L_0 - \mu_0 I_2 - 2
		\begin{pmatrix}
			\psi \\ -\chi
		\end{pmatrix}
		\begin{pmatrix}
			\psi & -\chi
		\end{pmatrix}.
	\end{equation*}

	Substituting the above relation into Eq.~\ref{sistema_inicial}, we can decompose the operator as:
	\begin{equation*}
		\begin{split}
			L & = \begin{pmatrix} 0 & L_0 \\ -L_1 & 0 \end{pmatrix}
			= -i \left[
				\begin{pmatrix}
					0      & i L_0 \\
					-i L_0 & 0
				\end{pmatrix}
				+ i
				\begin{pmatrix}
					0               & 0 \\
					\mu_0 + 2
					\begin{pmatrix}
						\psi \\ -\chi
					\end{pmatrix}
					\begin{pmatrix}
						\psi & -\chi
					\end{pmatrix} & 0
				\end{pmatrix}
				\right]
			= -i(\Gamma + B),
		\end{split}
	\end{equation*}
	where, $\Gamma$ is the self-adjoint operator
	\begin{equation*}
		\Gamma =
		\begin{pmatrix}
			0      & i L_0 \\
			-i L_0 & 0
		\end{pmatrix},
	\end{equation*}
	and $B$ is the following bounded, non-self-adjoint perturbation:
	\begin{equation*}
		B = i
		\begin{pmatrix}
			0               & 0 \\
			\mu_0 I_2 + 2
			\begin{pmatrix}
				\psi \\ -\chi
			\end{pmatrix}
			\begin{pmatrix}
				\psi & -\chi
			\end{pmatrix} & 0
		\end{pmatrix}.
	\end{equation*}

	Therefore, the spectrum of $L$ can be expressed in terms of the spectrum of $\Gamma + B$ as
	\begin{equation*}
		\sigma(L) = -i \, \sigma(iL) = -i \, \sigma(\Gamma + B).
	\end{equation*}

	Since $\Gamma$ is self-adjoint and $B$ is bounded, \textit{Theorem 2.1} of \citet{Cuenin_Tretter} applies directly to $\Gamma+B$. In particular, every spectral point of $\Gamma+B$ lies at distance at most $\|B\|_{L^2\mapsto L^2}$ from the real axis, hence
	\[
		\sigma(\Gamma + B) \subset \{ z \in \mathbb{C} : |\operatorname{Im}(z)| \leq \|B\|_{L^2 \mapsto L^2} \}.
	\]
	It remains to estimate $\|B\|_{L^2\mapsto L^2}$. Since $B$ is a multiplication operator by a matrix-valued function, its operator norm is bounded by the $L^\infty$ norm of that matrix, and therefore
	\begin{equation*}
		\|B\|_{L^2 \mapsto L^2}
		\leq |\mu_0| + 2 \left\| \begin{pmatrix} \psi \\ -\chi \end{pmatrix} \begin{pmatrix} \psi & -\chi \end{pmatrix} \right\|_\infty.
	\end{equation*}

	Substituting the explicit formulas for $\psi$ and $\chi$ from \eqref{soliton_profile}, we obtain $\|B\|_{L^2\mapsto L^2} \leq W$. Consequently,
	\[
		\sigma(\Gamma + B) \subset \{ z \in \mathbb{C} : |\operatorname{Im}(z)| \leq W \}.
	\]
	Using the relation $\sigma(L) = -i\,\sigma(\Gamma+B)$, we obtain \eqref{eq:banda}.
\end{proof}

While the previous result constrains the spectrum from \textit{infinity} (providing an upper bound on the real part), it is equally important to establish a lower bound on the modulus
of the eigenvalues, in order to control possible instabilities emerging from the origin. The following proposition provides such a bound, relating the eigenvalues
of $L$ to the spectra of the self-adjoint operators $L_0^2$ and $L_1^2$.

\begin{proposition}\label{prop_bola}
	For any eigenvalue $\Lambda \in \sigma_p(L)$, the following lower bound holds:
	\begin{equation*}
		|\Lambda|^2 \geq \min \big( \sigma(L_0^2) \cup \sigma(L_1^2) \big).
	\end{equation*}
\end{proposition}

\begin{proof}
	Let $\Lambda \in \sigma_p(L)$ and let $y \in H^1(\mathbb{R})^4$ be a corresponding normalized eigenfunction ($Ly = \Lambda y$, $\|y\|=1$).

	We calculate the squared norm directly
	\begin{equation*}
		\|Ly\|^2 = \langle Ly, Ly \rangle = |\Lambda|^2 \|y\|^2 = |\Lambda|^2.
	\end{equation*}
	Alternatively, using the adjoint operator, we write
	\begin{equation*}
		\langle Ly, Ly \rangle = \langle y, L^\dagger L y \rangle.
	\end{equation*}

	Using the fact that $$L^\dagger = \begin{pmatrix} 0 & -L_1 \\ L_0 & 0 \end{pmatrix},$$
	the product $L^\dagger L$ diagonalizes into a block structure
	\begin{equation*}
		L^\dagger L =
		\begin{pmatrix}
			0   & -L_1 \\
			L_0 & 0
		\end{pmatrix}
		\begin{pmatrix}
			0    & L_0 \\
			-L_1 & 0
		\end{pmatrix}
		=
		\begin{pmatrix}
			L_1^2 & 0     \\
			0     & L_0^2
		\end{pmatrix}.
	\end{equation*}

	Since $L_0^2$ and $L_1^2$ are positive self-adjoint operators, their spectra are bounded from below by the minimum of their respective spectral sets. Thus, for any normalized vector $y = (u,v)^T$, we have
	\begin{equation*}
		\langle y, L^\dagger L y \rangle
		= \langle u, L_1^2 u \rangle + \langle v, L_0^2 v \rangle
		\geq \min \big( \sigma(L_1^2) \cup \sigma(L_0^2) \big).
	\end{equation*}
	Combining the two expressions for $\|Ly\|^2$ yields the result.
\end{proof}

This estimate is particularly sharp when it is applied to the $H_{p.a.}$ subspace, where the translational zero mode is removed.

\begin{corollary}\label{corolario_bola}
	Consider the restricted operator $L_p : H_{p.a.} \mapsto H_{p.a.}$. Then:
	\begin{equation*}
		|\sigma_p(L_p)|^2 \geq \min \big( \sigma(L_{0,p}^2) \cup \sigma(L_{1,p}^2) \big)>0, \qquad\text{ for } \lambda \in (0, \lambda_c).
	\end{equation*}
\end{corollary}

\begin{proof}
	Since $L_0$ and $L_1$ are self-adjoint operators, we can use Lemma \ref{kerL1} and Corollary \ref{kerL0}, from where the result directly follows.
\end{proof}

\section{The non-relativistic limit $\lambda \to 0$}

Finally, to complete our spectral map, we analyze the behavior of the system in the non-relativistic limit $\lambda \to 0$. In this regime,
the spinor coupling weakens, and the system's stability properties should asymptotically approach those of the classical nonlinear Schrödinger (NLS) equation,
\citep{linear_instability_comech_guan_2014}. This limit serves as a validation argument for our analysis.

\begin{proposition}\label{spectrum_point_limit_Lp}
	There exists $\varepsilon > 0$ such that, for $\lambda \in (0,\varepsilon)$, the discrete spectrum of the restricted operator $L_p$ contains exactly two real eigenvalues with the asymptotic form:
	$$ \Lambda = \pm \sqrt{3(1-\Omega)} + O(\lambda). $$
\end{proposition}

\begin{proof}
	The proof proceeds in four steps. First, we square the eigenvalue problem to convert the non-self-adjoint first-order system into a second-order one. Second, we expand the squared operator for small $\lambda$. Third, we apply a Lyapunov--Schmidt reduction to extract a scalar equation governed by a P\"oschl--Teller potential. Fourth, we solve this reduced equation and translate the result back to the original spectral parameter.

	We restrict the operator $L$ to the positively alternating subspace $H_{p.a.}$. To decouple the parity components, we compose with the adjoint problem and define the squared operator
	\begin{equation}\label{eq_squared_nonrelativistic}
		\mathcal{M}_p(\lambda) \coloneqq \lambda^2 L_{0,p} L_{1,p},
	\end{equation}
	so that an eigenvalue $\Lambda$ of $L_p$ satisfies
	\begin{equation}\label{eq_squared_eigenvalue}
		\mathcal{M}_p(\lambda)\, y = -\lambda^2 \Lambda^2\, y.
	\end{equation}
	The rescaling factor $\lambda^2$ is introduced so that $\mathcal{M}_p(\lambda)$ admits a well-defined limit as $\lambda \to 0$. Since $L_{0,p}$ and $L_{1,p}$ are first-order operators with smooth bounded coefficients, $\mathcal{M}_p(\lambda)$ is a second-order operator with domain $H^2_{\mathrm{even}}(\mathbb{R}) \times H^2_{\mathrm{odd}}(\mathbb{R})$.

	Using the explicit soliton profiles \eqref{soliton_profile}, the potential terms admit the following expansions in $W^{1,\infty}(\mathbb{R})$:
	\begin{equation*}
		\psi(x) = \sqrt{2\lambda}\,\sech(x) + O(\lambda^{5/2}), \qquad
		\chi(x) = \sqrt{2}\,\lambda^{3/2}\sech(x)\tanh(x) + O(\lambda^{7/2}),
	\end{equation*}
	from which
	\begin{equation*}
		\tilde{N}(x) = 2\lambda\sech^2(x) + O(\lambda^3), \qquad
		2\psi(x)\chi(x) = 4\lambda^2\sech^2(x)\tanh(x) + O(\lambda^4).
	\end{equation*}
	Substituting these expressions into the definitions of $L_0$ and $L_1$ \eqref{def_operadores} and calculating the product $\lambda^2 L_{0,p} L_{1,p}$ yields
	\begin{equation}\label{eq_Mp_expansion}
		\mathcal{M}_p(\lambda) = \mathcal{M}_{-2} + \lambda\,\mathcal{M}_{-1} + \lambda^2\,\mathcal{M}_0 + \lambda^3\,\mathcal{R}(\lambda),
	\end{equation}
	where $\sup_{0<\lambda<\lambda_c}\|\mathcal{R}(\lambda)\| < \infty$, and
	\begin{equation*}
		\mathcal{M}_{-2} = \begin{pmatrix} 0 & 0 \\ 0 & \Omega I \end{pmatrix}, \qquad
		\mathcal{M}_{-1} = \Omega \begin{pmatrix} 0 & -\partial_x \\ \partial_x & 0 \end{pmatrix},
	\end{equation*}
	\begin{equation*}
		\mathcal{M}_0 =
		\begin{pmatrix}
			-\partial_x^2 + (1-\Omega)(1 - 6\sech^2(x)) & 0                                           \\
			0                                           & -\partial_x^2 + (1+\Omega)(1 - 2\sech^2(x))
		\end{pmatrix}.
	\end{equation*}
	This is the one-dimensional, parity-restricted analogue of formulas (5.9)--(5.12) in \citet{linear_instability_comech_guan_2014}.

	We introduce $\Sigma \coloneqq \Lambda^2$ and rewrite \eqref{eq_squared_eigenvalue} in block form:
	\begin{equation}\label{sis_norel_red}
		\begin{pmatrix}
			\mathcal{M}_p^{11}(\lambda) + \lambda^2 \Sigma & \mathcal{M}_p^{12}(\lambda)                    \\
			\mathcal{M}_p^{21}(\lambda)                    & \mathcal{M}_p^{22}(\lambda) + \lambda^2 \Sigma
		\end{pmatrix}
		\begin{pmatrix} u \\ v \end{pmatrix} = 0.
	\end{equation}
	From \eqref{eq_Mp_expansion}, the $(2,2)$-block satisfies $\mathcal{M}_p^{22}(\lambda) + \lambda^2 \Sigma = \Omega I + O(\lambda)$. Since $\Omega > 0$, this block is invertible for $\lambda$ sufficiently small, and the second row of \eqref{sis_norel_red} can be solved for $v$ as a function of $u$:
	\begin{equation*}
		v = -\bigl(\mathcal{M}_p^{22}(\lambda) + \lambda^2 \Sigma\bigr)^{-1}\,\mathcal{M}_p^{21}(\lambda)\, u = -\lambda\,\partial_x u + O(\lambda^2)\,u.
	\end{equation*}

	Recall that the $(2,2)$-block might not be invertible if $\Sigma = -\Omega/\lambda^2+O(1/\lambda)$. However, this regime is not of interest for our instability analysis because the corresponding eigenvalues $\Lambda = \pm i\sqrt{\Omega}/\lambda + O(1/\sqrt{\lambda})$ do not lie on the real axis (for sufficiently small $\lambda$) and thus do not contribute to spectral instability. In fact, for other values of $\Sigma$, it is satisfied that $|\Sigma(\lambda = 0)| < \infty.$

	Substituting the expression for $v$ back into the first row and dividing by $\lambda^2$ eliminates the odd component entirely, yielding a \emph{scalar} equation for the even function $u$ alone:
	\begin{equation}\label{eq_reduced_even_problem}
		\mathcal{S}(\lambda, \Sigma)\, u = 0,
	\end{equation}
	where
	\begin{equation}\label{eq_reduced_even_operator}
		\mathcal{S}(\lambda, \Sigma) = \mathcal{H}_+ + \frac{\Sigma}{1-\Omega} + \mathcal{E}(\lambda, \Sigma),
	\end{equation}
	with $\mathcal{H}_+ \coloneqq -\partial_x^2 + 1 - 6\sech^2(x)$ and $\|\mathcal{E}(\lambda,\Sigma)\|_{H^2 \mapsto L^2} \le C\lambda$. The operator $\mathcal{S}$ acts on $H^2_{\mathrm{even}}(\mathbb{R})$, and the reduction is the one-dimensional, parity-restricted version of the Rayleigh--Schr\"odinger scheme in \citet{linear_instability_comech_guan_2014} (see also Chapters~VII and VIII of \citet{kato}).

	The key observation is that the leading-order operator $\mathcal{H}_+$ is a 
	Schrödinger operator with the \emph{P\"oschl--Teller} potential, whose spectrum is explicitly known (see e.g.\ \citep{raban:2022}). On the full line, $\mathcal{H}_+$ has exactly two discrete eigenvalues: $-3$ (with even eigenfunction $u_0(x) = \sech^2(x)$) and $0$ (with odd eigenfunction). Restricting to the even subspace $L^2_{\mathrm{even}}(\mathbb{R})$ removes the zero eigenvalue, so that $-3$ is the \emph{unique} discrete eigenvalue of $\mathcal{H}_+|_{L^2_{\mathrm{even}}}$.

	Setting $\lambda = 0$ in \eqref{eq_reduced_even_operator}, we find that $\mathcal{S}(0, \Sigma)$ has a one-dimensional kernel if and only if $\Sigma/(1-\Omega) = 3$, i.e.,
	\begin{equation*}
		\Sigma_0 = 3(1-\Omega).
	\end{equation*}

	To track how this kernel deforms for $\lambda > 0$, we apply a standard Lyapunov--Schmidt decomposition. Let $P$ denote the orthogonal projection onto $\Span\{u_0\}$ and write $u = u_0 + w$ with $w \perp u_0$. Since $-3$ is an isolated simple eigenvalue, the operator $\mathcal{S}(0, \Sigma_0)|_{(I-P)L^2_{\mathrm{even}}}$ is invertible. By the implicit function theorem, for $(\lambda, \Sigma)$ near $(0, \Sigma_0)$ there exists a unique smooth correction $w = w(\lambda, \Sigma)$ solving the projected equation $(I - P)\,\mathcal{S}(\lambda, \Sigma)(u_0 + w) = 0$. Substituting back and projecting onto $\Span\{u_0\}$ yields the scalar \emph{bifurcation equation}
	\begin{equation*}
		F(\lambda, \Sigma) \coloneqq \bigl\langle u_0,\; \mathcal{S}(\lambda, \Sigma)\,(u_0 + w(\lambda, \Sigma)) \bigr\rangle = 0.
	\end{equation*}
	Straightforward calculations show that $F(0, \Sigma_0) = 0$ and
	\begin{equation*}
		\partial_\Sigma F(0, \Sigma_0) = \frac{1}{1-\Omega}\,\|u_0\|_2^2 \neq 0.
	\end{equation*}
	Therefore, the implicit function theorem provides a unique smooth branch $\Sigma = \Sigma(\lambda)$ with
	\begin{equation}\label{eq_sigma_small_lambda}
		\Sigma(\lambda) = 3(1-\Omega) + O(\lambda).
	\end{equation}

	Since $\mathcal{H}_+$ has Morse index one on $L^2_{\mathrm{even}}(\mathbb{R})$ and its remaining spectrum is bounded below by a positive gap, the Riesz spectral projection associated with $-3$ has rank one and persists for $0 < \lambda < \varepsilon$ by the stability of isolated spectral subspaces (Chapter~IV of \citet{kato}). Consequently, \eqref{eq_sigma_small_lambda} is the \emph{only} positive branch of the reduced problem.

	Returning to the original spectral parameter through $\Lambda^2 = \Sigma(\lambda)$, we conclude that $\mathcal{M}_p(\lambda)$ has exactly one negative eigenvalue $-\lambda^2 \Sigma(\lambda)$, and the operator $L_p$ possesses exactly one pair of real eigenvalues:
	\begin{equation*}
		\Lambda_\pm(\lambda) = \pm\sqrt{\Sigma(\lambda)} = \pm\sqrt{3(1-\Omega)} + O(\lambda). \qedhere
	\end{equation*}
\end{proof}

This result confirms that the instability mechanism persists in the non-relativistic limit and aligns analytically with the instability rates observed numerically in \citep{chaos_nlde}.

\section{Instability in the subcritical region}
Having established the existence of a positive eigenvalue in the non-relativistic limit $\lambda \to 0$ (Proposition~\ref{spectrum_point_limit_Lp}) and determined the global
bounds on the spectrum (Proposition~\ref{prop:banda}), we now proceed to the main result of this work. We combine these local and global properties to demonstrate that the
instability persists throughout the entire subcritical regime. By using the continuity of the spectrum and the topological properties of the eigenvalue flow, we show that the unstable mode identified near $\lambda=0$ cannot vanish or stabilize until the critical threshold $\lambda_c$ is reached.
\begin{theorem}
	The soliton $\Psi_-$ of Eq.~\eqref{NLDE} is unstable for all $\lambda \in (0, \lambda_c)$.
\end{theorem}

\begin{proof}
	The proof relies on a topological continuity argument. We track the movement of eigenvalues in the complex plane as the parameter $\lambda$ varies.

	Recall from Corollary~\ref{corolario_bola} that for the restricted operator $L_p$, the non-zero spectrum is strictly bounded away from the origin. We define the spectral gap function:
	\begin{equation*}
		R(\lambda) := \min \big( \sigma(L_{0, p}^2) \cup \sigma(L_{1, p}^2) \big)^{1/2}.
	\end{equation*}
	By the spectral properties of the constituent operators (Lemma \ref{kerL1} and Corollary \ref{kerL0}), $R(\lambda)$ is continuous and strictly positive for $\lambda \in (0, \lambda_c)$.

	We consider the compact interval $I = [ \varepsilon/2, \lambda_c - \delta/2]$, where $\varepsilon$ and $\delta$ are small constants chosen such that the asymptotic results of Proposition~\ref{spectrum_point_limit_Lp} and Theorem~\ref{teorema_bifurcacion} hold at the boundaries. We define the global bounds over this interval:
	\begin{equation*}
		R_{\min} := \min\{\min_{\lambda \in I} \{R(\lambda), \nu_1\} \} > 0, \qquad
		W_{\max} := \max_{\lambda \in I} W(\lambda),
	\end{equation*}
	where $W(\lambda)$ is the spectral bound from Proposition~\ref{prop:banda} and $\nu_1 = \sqrt{1-(1+\lambda^2)\Omega}$, limits the essential spectrum (Proposition~\ref{espectro_essencial}).

	With these bounds, we construct a rectangular contour $B$ in the right half of the complex plane that strictly encloses the region of potential instability, avoiding the origin and the essential spectrum (see Fig.~\ref{box_fig}):
	\begin{equation*}
		B := \{x+iy \in \mathbb{C} \,:\, 0\leq x-\tfrac{\sqrt{3}}{4} R_{\min} \leq W_{\max}, \quad |y| \leq R_{\min}/4\}.
	\end{equation*}

	\begin{figure}[h]
		\centering
		\begin{tikzpicture}[scale=1.5]
			\usetikzlibrary{patterns}

			\def\Gap{1.0}      
			\def\BoxW{3.5}     
			\def\BoxH{0.6}     

			\draw[->, thick, gray] (-1.5,0) -- (5,0) node[black, right] {$\text{Re}(\Lambda)$};
			\draw[->, thick, gray] (0,-1.5) -- (0,1.5) node[black, above] {$\text{Im}(\Lambda)$};

			\draw[dashed, red, thick] (0,0) circle (\Gap);
			\node[red, anchor=south west] at (-1.25, 1.0) {$|\Lambda| = R_{\min}$};
			\fill[white] (0,0) circle (\Gap); 
			\draw[dashed, red] (0,0) circle (\Gap);
			\node at (0,0) {$\times$}; 
			\node[anchor=west] at (0.01,-0.12) {$0$};
									
			\node[anchor=west] at (0.15,0.25) {$\ell$};			
			\node[anchor=west] at (3.5,0.25) {${r}$};

			\draw[thick, blue, pattern=north east lines, pattern color=blue!20]
			(\Gap/2, -\BoxH) rectangle (\BoxW, \BoxH);

			\node[blue] at (\BoxW-0.5, \BoxH) [anchor=south east] {Box $B$};
			\draw[blue, thick] (\Gap/2, -\BoxH) -- (\Gap/2, \BoxH);

			\draw[dashed, thick] (\BoxW-0.5, -1.5) -- (\BoxW-0.5, 1.5);
			\node[above] at (\BoxW-0.5,1.5) {$\text{Re}(\Lambda)=W_{\max}$};
		\end{tikzpicture}
		\caption{Schematic representation of the counting box $B$ in the complex plane. The spectral gap $R_{\min}$ (dashed circle) ensures that no non-zero eigenvalues can cross the imaginary axis or the origin to exit the box.}
		\label{box_fig}
	\end{figure}
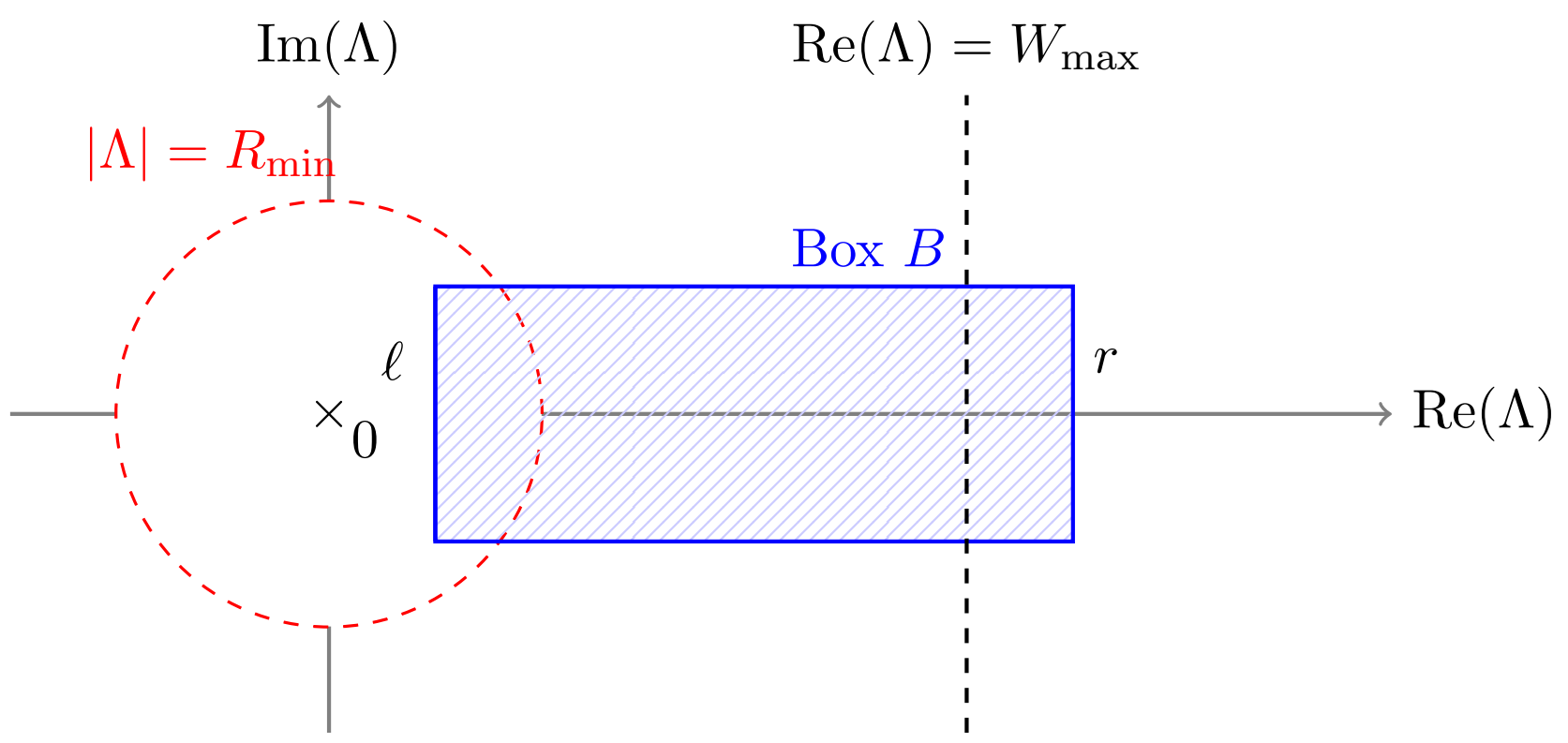

	Let $N_B(\lambda)$ be the number of eigenvalues of $L$ contained within $B$ (counted with multiplicity). The boundary of $B$ consists of four segments:
	\begin{enumerate}
		\item \textit{Left edge} $\ell$ ($\text{Re}(\Lambda) = \sqrt{3}/4 \, R_{\min}$): No eigenvalues can cross $\ell$ because $|\Lambda| \geq R_{\min}$ for the point spectrum, according to Corollary~\ref{corolario_bola}.
		\item \textit{Right edge} $r$ ($\text{Re}(\Lambda) = W_{\max} + \sqrt{3}/4 \, R_{\min}$): No eigenvalues belong to $r$ due to the a priori bound in Proposition~\ref{prop:banda}.
		\item \textit{Top/Bottom edges}: Crucially, because $L$ is a real operator with an off-diagonal block structure, its nonreal eigenvalues occur in quartets $(\Lambda, -{\Lambda}^\star, -\Lambda, {\Lambda}^\star)$. Therefore, any eigenvalues entering or leaving $B$ across the horizontal edges must do so in conjugate pairs, meaning the number of eigenvalues changes by an even integer.
	\end{enumerate}

	Because eigenvalues can only enter or leave the box in pairs, the parity of the number of eigenvalues inside $B$ is a topological invariant with respect to $\lambda$:
	\begin{equation*}
		N_B(\lambda) \equiv N_B(\varepsilon/2) \pmod 2 \qquad \text{for all } \lambda \in I.
	\end{equation*}

	From Proposition~\ref{spectrum_point_limit_Lp}, in the non-relativistic limit $\lambda \to 0$, the right half-plane contains exactly one simple real eigenvalue. Therefore, at $\lambda = \varepsilon/2$, we have:
	\begin{equation*}
		N_B(\varepsilon/2) = 1.
	\end{equation*}

	Consequently, $N_B(\lambda)$ must be an odd integer (and thus at least 1) for all $\lambda \in I$. This proves the existence of at least one eigenvalue with a strictly positive real part across the entire subcritical range $\lambda \in (0, \lambda_c)$, rigorously confirming the instability of the soliton $\Psi_-$.
\end{proof}

\section{Conclusion}

In this work, we have provided a rigorous analytical characterization of the spectral stability of solitary wave solutions to the nonlinear Dirac equation under dissipation and parametric forcing in the subcritical regime. By combining asymptotic analysis with spectral bounds, we have confirmed the instability mechanisms previously observed in numerical simulations.

Our main results can be summarized as follows:

\begin{enumerate}

	\item \textit{Bifurcation at criticality:} We performed a local spectral analysis around the critical parameter $\lambda_c$. Using a Lyapunov--Schmidt reduction based on the critical Jordan chain of $L_p$ (Theorem~\ref{teorema_bifurcacion}), we derived the explicit asymptotic law $\Lambda(\lambda) \propto \pm i \sqrt{\lambda - \lambda_c}$. This result analytically confirms the mechanism of instability: a bifurcation where a pair of real unstable eigenvalues collide at the origin and transition onto the imaginary axis, providing a precise local verification of the Vakhitov-Kolokolov criterion.

	\item \textit{Spectral enclosure:} We established a priori bounds on the location of the spectrum in the complex plane (Proposition~\ref{prop:banda}). Furthermore, we proved that the spectrum of $L_p$ lies outside the spectral ball of radius $R(\lambda)$ centered at the origin, a crucial property that allows the isolation of discrete instability modes.

	\item \textit{The non-relativistic limit:} In the limit $\lambda \to 0$, we squared the spectral problem and reduced it, via a Lyapunov--Schmidt argument, to a P\"oschl--Teller Schr\"odinger operator. This yields a unique real pair $\Lambda = \pm\sqrt{3(1-\Omega)} + O(\lambda)$ in the positive alternating subspace and identifies the non-relativistic origin of the instability.

	\item \textit{Global instability via topology:} Using a topological continuity argument, we extended the local result from the limit $\lambda \to 0$ to the entire subcritical interval $(0, \lambda_c)$. By defining a contour that avoids the spectral gap and the essential spectrum, we showed that the parity of the eigenvalue count in a box is invariant. Since the system possesses an odd number of unstable eigenvalues near $\lambda=0$, the soliton must remain unstable for all $\lambda < \lambda_c$.
\end{enumerate}

These findings provide a mathematical foundation for the Vakhitov-Kolokolov stability criterion in the context of spinor fields, \citep{comechVK}. While the Vakhitov-Kolokolov slope condition predicts a transition from instability to stability at $\lambda_c$, our work rigorously proves the global existence of instability throughout the entire subcritical region.

\textit{Future work:} The natural continuation of this research is to rigorously establish the conditions for stability in the supercritical regime $\lambda > \lambda_c$. While the absence of unstable eigenvalues is a necessary condition for stability, it is not sufficient. As shown numerically in \citep{chaos_nlde}, the supercritical regime exhibits a richer phenomenology, including both stable and unstable regions depending on the applied dissipation and parametric forcing.

\section*{Statements and Declarations}

\textbf{Funding:} This work was supported by FPU24/00951 (Ministerio de Ciencia, Innovaci\'on y Universidades, Spain), by FQM-415 (Plan Propio, Universidad de Sevilla, Spain), and by the IMUS Mar\'ia de Maeztu grant CEX2024-001517-M (Apoyo a Unidades de Excelencia Mar\'ia de Maeztu), funded by MICIU/AEI/10.13039/501100011033.

\textbf{Competing interests:} The author has no relevant financial or non-financial interests to disclose.

\section*{Acknowledgments}
The author thanks Niurka R. Quintero and Renato \'Alvarez-Nodarse for their valuable discussions and insightful comments.
\bibliographystyle{plainnat}
\bibliography{subcritical_instability_nlde} 

\end{document}